\documentclass[a4paper,twosides]{article}
\usepackage{CJK,multicol,multirow,graphics,fancyhdr,epstopdf}
\usepackage{amsmath,amsfonts,amssymb,bm,upgreek,mathrsfs,ccmap,mathcomp,dsfont,makecell}
\usepackage[pagewise,switch,columnwise]{lineno}
\usepackage[compress,nospace]{cite}
\usepackage[dvipsnames]{xcolor}
\usepackage{wasysym}

\usepackage[colorlinks,linkcolor=blue,urlcolor=blue,citecolor=blue,anchorcolor=blue,pdfborder=000,
  pdftitle={Changepdftitle}, pdfauthor={Changepdfauthor},
  pdfcreator=CHIN. PHYS. LETT., pdfproducer=CPL]{hyperref}

\newcommand{\ucite}[1]{$\!^\text{\cite{#1}}$} 
\definecolor{blue}{RGB}{46,48,146}

\def\headrule{\kern 1mm \hrule width 17cm \kern -1mm}%
\def\footnoterule{\kern 1mm \hrule width 7cm \kern 2.2mm}%
\newcommand{\figcaption}[3]{\centerline{\footnotesize \begin{tabular}{p{#1cm}}{\bf Fig.~{#2}. }#3\end{tabular}}} 

\newcommand{\ket}[1]{|#1\rangle}

\newcommand{\fref}[2]{\hyperlink{f#1}{#2}} 
\newcommand{\fl}[1]{\hypertarget{f#1}{}} 

\newcommand{\cplyear}{2025} \newcommand{\cplvol}{42}
\newcommand{\cplno}{9} \newcommand{\cplpagenumber}{090706}
 \newcommand{\cplpage}{\cplpagenumber-\thepage}

\begin{document}

\begin{CJK}{GBK}{song}\vspace* {-4mm} \begin{center}
\large\bf{\boldmath{Floquet Non-Abelian Topological Charges and Edge States}}
\footnotetext{\footnotesize\hspace*{-5.4mm}
\noindent$^{*}$Corresponding author. Email: \href{mailto:zhoulw13@u.nus.edu}{zhoulw13@u.nus.edu}

\noindent\copyright\,{\cplyear}
\href{http://www.cps-net.org.cn}{Chinese Physical Society} and \href{http://www.iop.org}{IOP Publishing Ltd}. All rights, including for text and data mining, AI training, and similar technologies, are reserved.}
\\[3mm]
\normalsize \rm{} Jiaxin Pan$^{1}$ and Longwen Zhou$^{1,2,3*}$
\\[3mm]\small\sl $^{1}$College of Physics and Optoelectronic Engineering, Ocean University of China, Qingdao 266100, China

$^{2}$Key Laboratory of Optics and Optoelectronics, Qingdao 266100, China

$^{3}$Engineering Research Center of Advanced Marine Physical Instruments and Equipment of MOE,\\ Qingdao 266100, China
\\[3mm]\normalsize\rm{}(Received 3 April 2025; accepted manuscript online 21 July 2025)
\end{center}
\end{CJK}
\vskip 1.5mm

\small{\narrower Non-Abelian topological insulators are characterized by matrix-valued, non-commuting topological charges with regard to more than one energy gap. Their descriptions go beyond the conventional topological band theory, in which an additive integer like the winding or Chern number is endowed separately with each (degenerate group of) energy band(s). In this work, we reveal that Floquet (time-periodic) driving could not only enrich the topology and phase transitions of non-Abelian topological matter, but also induce bulk-edge correspondence unique to nonequilibrium setups. Using a one-dimensional, three-band model as an illustrative example, we demonstrate that Floquet driving could reshuffle the phase diagram of the non-driven system, yielding both gapped and gapless Floquet band structures with non-Abelian topological charges. Moreover, by dynamically tuning the anomalous Floquet $\pi$-quasienergy gap, non-Abelian topological transitions inaccessible to static systems could arise, leading to much more complicated relations between non-Abelian topological charges and Floquet edge states. These discoveries put forth periodic driving as a powerful scheme of engineering non-Abelian topological phases and incubating unique non-Abelian band topology beyond equilibrium.

\par}\vskip 3mm
\noindent{{DOI: \href{http://dx.doi.org/10.1088/0256-307X/\cplvol/\cplno/\cplpagenumber}{10.1088/0256-307X/\cplvol/\cplno/\cplpagenumber}}
\hspace*{2.56cm}
\noindent{\narrower{CSTR:
\href{https://cstr.cn/32039.14.0256-307X.\cplvol.\cplno.\cplpagenumber}{32039.14.0256-307X.\cplvol.\cplno.\cplpagenumber}}

\par}\vskip 5mm}
\begin{multicols}{2}
\baselineskip=12pt plus.2pt minus.2pt

{\it 1. Introduction.}
Topology has emerged as a new organizing principle of condensed matter and phase transitions in the last decades. By integrating conventional band theory with topological classifiers, a large group of insulators, superconductors, and semimetals can be characterized by the topological invariants of their bulk energy bands.\ucite{1,2,3,4,5} These invariants, with the Chern and winding numbers being representatives, are integer quantized, additive, and pertaining to Abelian groups. Recently, a new class of topological matter was discovered, whose multigap band topology is yet associated with non-Abelian topological charges (NATCs) like quaternions. These non-commuting charges may originate from the eigenframe rotation of ${\cal PT}$- or ${\cal C}_{2}{\cal T}$-symmetric Hamiltonians with gapped band structures,\ucite{6,7,8,9,10} whose eigenvectors are real and do not produce geometric phase windings in momentum space. In gapless systems, NATCs can also depict the braiding of nodal points and nodal lines in multiband topological  semimetals.\ucite{11,12,13,14,15,16,17,18} These intriguing discoveries have not only promoted the development of topological band theory with non-Abelian characters, but also inspired great progress in the experimental realization of non-Abelian topological matter with photonic,\ucite{6,7,8}  acoustic,\ucite{19, 20, 21,22,23,24} electronic,\ucite{25,26} ionic,\ucite{27,28} and cold-atom\ucite{29} setups, yielding potential applications in channel multiplexing and path-dependent topological mode converters.\ucite{30,31}

Floquet driving offers a successful route to engineer Bloch bands and create nonequilibrium topological states.\ucite{32,33,34,35,36,37,38,39,40,41} High-frequency drives may induce band-inversion and produce topological gaps at Dirac points.\ucite{42,43,44,45} Near-resonant drivings may generate long-range coupling and result in Floquet phases with large topological invariants and many edge modes.\ucite{46,47,48,49} Stroboscopically, the dynamics of a driven system is governed by its Floquet operator $U=\mathsf{\hat T}{\rm e}^{-\frac{{\rm i}}{\hbar}\int_{0}^{T}H(t){\rm d}t}$, where $\mathsf{\hat T}$ performs time-ordering and $H(t)=H(t+T)$ is the time-periodic Hamiltonian. As the spectrum ${\rm e}^{-{\rm i}E}$ of $U$ is defined on a ${\rm U}(1)$ loop, any adjacent pair of Floquet bands may touch and yield a topological transition. Anomalous Floquet phases could then emerge, which are featured by degenerate edge modes at $E=\pm\pi$ or chiral Floquet bands weaving around the whole quasienergy Brillouin zone (BZ) $E\in[-\pi,\pi]$.\ucite{50,51,52,53} These intriguing phases are yet captured by Abelian topological charges, in analogy to static insulators or semimetals.\ucite{54,55,56,57} Very recently, it was found that Floquet driving may endow multigap systems with non-Abelian topology.\ucite{58,59,60} However, many important issues are still open. First, an overall phase diagram for Floquet non-Abelian topological phases (NATPs) is missing even for simple tight-binding models. Second, NATCs are usually related to multigap insulating bands, which could be drastically reshaped by Floquet driving and form indirect overlaps in momentum space, making the existing non-Abelian characterizations questionable. Third, the non-commuting nature of NATCs allows path-dependent transitions between different non-Abelian phases, yielding contentious multifold correspondence\ucite{61} between NATCs and Floquet edge modes.

\end{multicols}
\vskip -1mm

\fl{1}\centerline{\includegraphics{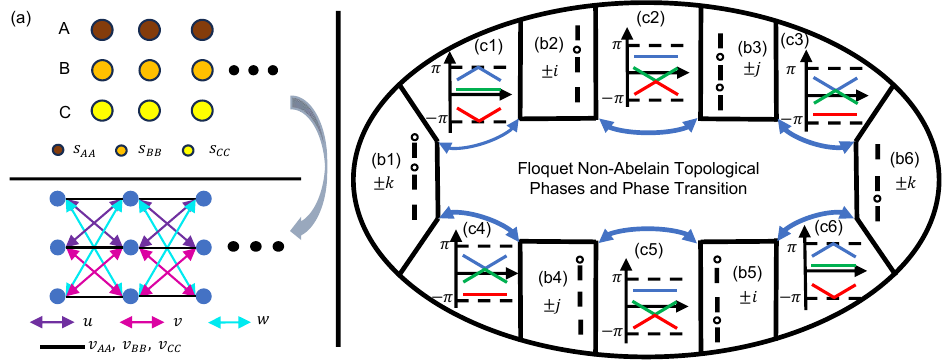}}
\vskip 1mm
\figcaption{15}{1}{The Floquet model and its topology. (a) The top (bottom) panel is the model in the first (second) half of a driving period. Brown, orange and yellow dots are sublattices A, B and C with different onsite potentials $s_{AA}$, $s_{BB}$ and $s_{CC}$. Black lines are hoppings between the same sublattice in nearest-neighbor~(NN) unit cells. Purple, pink, and cyan arrows are hoppings between different sublattices in NN unit cells. (b1)--(b6) Edge states characterized by different NATCs under open boundary conditions. Black circles and lines denote edge and bulk states. (c1)--(c6) Quasienergy spectra at critical points between different phases. The first, second and third bands are marked by red, green and blue lines, respectively. Boundaries of the quasienergy BZ are labeled by black dashed lines.}

\begin{multicols}{2}
\baselineskip=12pt plus.2pt minus.2pt

To address these issues, we investigate a one-dimensional (1D) model with three bands and subject to time-periodic quenches. Via quaternion charges, we build the phase diagram and identify rich varieties of driving-induced non-Abelian topological insulators (NATIs) and semimetals with the latter not reported before. Strikingly, the tunable Floquet gap at quasienergy $\pi$ creates a new transition path between distinct NATPs [see Figs.\,\fref{fig1}{1(c1)}--\fref{fig1}{1(c6)}] and a unique zone for Floquet topological edge modes to appear [see Figs.\,\fref{fig1}{1(b1)}--\fref{fig1}{1(b6)}]. Theoretical rules of Floquet non-Abelian bulk-edge correspondence are further obtained.

{\it 2. Model and Theory.}
In momentum space $k_x\in[-\pi,\pi]$, the time-periodic Hamiltonian of our Floquet system takes the form (with $\hbar=1$ and $T=2$)
\begin{equation}\label{Floquet Hamiltonian}
	H(t)=\begin{cases}
		H_{1}(k_{x}), & t\in[\ell,\ell+1),\\
		H_{2}(k_{x}), & t\in[\ell+1,\ell+2),		
	\end{cases}
\end{equation}
where $\ell\in{\mathbb Z}$, $H_1(k_x)={\rm diag}(s_{AA},s_{BB},s_{CC})$, and
\begin{align}
	\label{H2}H_{2}(k_{x}) & = 2 \left[\begin{array}{ccc}
	v_{AA}\cos{k_{x}}  &   u\sin{k_{x}}   &  w\sin{k_{x}}  \\
       u\sin{k_{x}}    & v_{BB}\cos{k_{x}}&  v\sin{k_{x}} \\
	   w\sin{k_{x}}    &   v\sin{k_{x}}   &v_{CC}\cos{k_{x}}
	\end{array}\right].
\end{align}
The elements $s_{AA}$, $s_{BB}$, and $s_{CC}$ denote onsite potentials of sublattices A, B, and C; $v_{AA}$, $v_{BB}$, and $v_{CC}$ denote hopping amplitudes between nearest-neighbor~(NN) unit cells of the same sublattice, respectively. $u={\rm i}v_{AB}={\rm i}v_{BA}$, $v={\rm i}v_{BC}={\rm i}v_{CB}$, and $w={\rm i}v_{AC}={\rm i}v_{CA}$ are hopping amplitudes between NN unit cells of different sublattices [see Fig.\,\fref{fig1}{1(a)}]. In our model, $H_{1}(k_{x})$ and $H_{2}(k_{x})$ are both real. The resulting Floquet operator is $U={\rm e}^{-{\rm i}H_{2}(k_{x})}{\rm e}^{-{\rm i}H_{1}(k_{x})}$, whose quasienergies $E_{n}$ and eigenstates $\ket{\phi_{n}}$ ($n=1,2,3$) are obtained by solving the eigenvalue equation $U\ket{\phi_{n}}={\rm e}^{-{\rm i}E_{n}}\ket{\phi_{n}}$. When $\left[H_{1}(k_{x}),H_{2}(k_{x})\right]\neq 0$, the Floquet operator $U$ lacks $\mathcal{PT}$-symmetry and its eigenstates are complex, making the non-Abelian characterization unavailable. This issue can be resolved by transforming $U$ into a symmetric time frame,\ucite{62} yielding $U'={\rm e}^{-{\rm i}H_{2}/2}{\rm e}^{-{\rm i}H_{1}}{\rm e}^{-{\rm i}H_{2}/2}$. $U'$ shares the same spectrum with $U$ due to their unitary equivalence.
The eigenstates of $U'$ are yet real vectors, as the associated effective Hamiltonian $H'_{\rm eff}={\rm i}\ln{U'}/2$ is real and ${\cal PT}$ symmetric. The fundamental homotopy group of $H'_{\rm eff}$ is $\pi_1[{\rm O}(3)/{\rm O}(1)^3]=Q$. $Q= (1, \pm i, \pm j, \pm k, -1)$ forms the non-Abelian quaternion group, which has three imaginary units ${i, j, k}$ with $ij=k$, $jk=i$, $ki=j$, and $i^2=j^2=k^2=-1$\ucite{6} [for details, see Secs.~I and II of the Supplemental Material (SM)]. The eigenvectors $(\ket{u_{1}},\ket{u_{2}},\ket{u_{3}})$ of $H'_{\rm eff}$ define a non-Abelian Berry connection with matrix elements $\left[A(k_x)\right]_{mn}=\langle{u_{m}|\partial_{k_x}|u_{n}\rangle}$ ($m,n=1,2,3$) in $k_x$-space, whose SO(3) representation reads $A(k_x)=\sum_{n=1}^{3}\beta_{n}(k_x)L_{n}$. Here, $(L_{n})_{lm}=-\epsilon_{nlm}$, and $\epsilon_{nlm}$ is the Levi--Civita symbol. With the same coefficients $\beta_{n}$, $A(k_x)$ has an SU(2) representation $\bar{A}(k_x)=\sum_{n=1}^{3}\beta_{n}t_{n}$, where $t_{n}=-{\rm i}\sigma_{n}/2$ and $\sigma_{1,2,3}$ are Pauli matrices $\sigma_{x,y,z}$, respectively. The NATCs of our system is then defined as $q={\hat{\mathsf P}}{\rm e}^{\oint_{\rm BZ} \bar{A}(k_x){\rm d}k_x}$, where ${\hat{\mathsf P}}$ performs the path-ordering from $k_x=-\pi\rightarrow\pi$. Each of the matrix-valued charges can be represented by a quaternion as $1\leftrightarrow\sigma_{0}$, $i\leftrightarrow-{\rm i}\sigma_{1}$, $j\leftrightarrow-{\rm i}\sigma_{2}$ and $k\leftrightarrow-{\rm i}\sigma_{3}$. The quaternion group $Q$ forms the NATCs characterizing the topological phases of our Floquet system.

\emph{\it 3. Floquet NATPs.}
Transitions among distinct Floquet NATPs are accompanied by the closing/reopening of direct gaps between quasienergy bands. We thus identify the phase boundaries of Floquet NATPs via the function $\Delta_{mn}\equiv\min_{k_{x}\in[-\pi,\pi)}|E_{m}-E_{n}|$ in parameter space, where $m,n=1,2,3$ and $m<n$. A phase transition occurs when $\Delta_{mn}=0$ for a given pair $(E_{m}, E_{n})$. Note that the case $\Delta_{13}=0$ relates to the touch of Floquet bands around $E=\pm\pi$, which yields a unique path of topological transition that is forbidden in static systems. With all phase boundaries identified and the NATCs computed, we establish the phase diagram of Floquet NATPs as shown in Fig.\,\fref{fig2}2, which presents several key features.

\vskip 4mm

\fl{2}\centerline{\includegraphics{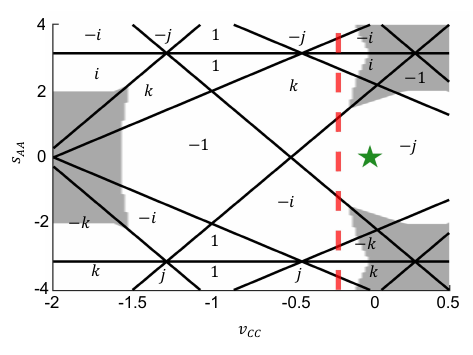}}
\vskip 1mm
\figcaption{7.8}{2}{Phase diagram of Floquet NATPs. Black lines are phase boundaries determined by the gap functions. The green star is the base point to define the order of quasienergy bands. Gray regions denote semimetal phases with direct gaps, indirect band overlaps and well-defined NATCs. We set $s_{AA}=s_{CC}$, $v_{CC}=-1-v_{BB}$, $v_{AA}=u=-v=1$ and $s_{BB}=w=0$.}

\vskip 4mm

First, all possible NATCs in the quaternion group $Q$ and their mutual transitions are found in Fig.\,\fref{fig2}2, while only part of them could survive in the same parameter space of the associated static model (see Fig.\,S3 of the SM). Moreover, certain trivial phases with quaternion charge $1$ in the static setting are transformed to Floquet NATPs via our simple quenching protocol (see also Fig.\,S3 of the SM). Therefore, Floquet drivings applied to a static system could flexibly induce and greatly enrich the NATPs and phase transitions therein, offering an efficient means of realizing and engineering non-Abelian topological matter beyond equilibrium.

Second, the white and gray regions in Fig.\,\fref{fig2}2 correspond to Floquet NATIs and non-Abelian topological semimetals (NATSs), respectively. Although the former was reported before, the latter has not been identified due to the lack of a systematic phase diagram in previous studies. These NATSs are gapless due to the indirect overlap of their Floquet bands in momentum space. This makes them different from Dirac or Weyl semimetals, which are depicted by Abelian topological charges like $\pi$-Berry phases or Chern numbers at band-touching points. With separable bands, Floquet NATSs can still be characterized by quaternions, whose values are shown in Fig.\,\fref{fig2}2. Meanwhile, the edge states associated with NATCs could survive in the gap(s) where the Floquet bands of the corresponding NATSs have no overlaps (see Sec.\,IV of the SM for more details).
Our discoveries thus reveal a unique type of Floquet topological semimetal with a non-Abelian
origin.

Third, periodic drivings enable a transition path between distinct NATPs that is unique to Floquet systems. It is achieved by tuning the gap between the quasienergy bands around $E=\pm\pi$. This process is illustrated in Figs.\,\fref{fig3}{3(a)}--\fref{fig3}{3(c)}, which is related to the change of system parameter $s_{AA}$ along the red dashed line in Fig.\,\fref{fig2}2. Starting at $s_{AA}=2$ with the quaternion charge $k$ and the spectrum in Fig.\,\fref{fig3}{3(a)}, we can raise $s_{AA}$ to induce a transition from charge $k$ to $i$ [Fig.\,\fref{fig3}{3(c)}] via closing the gap at $E=\pm\pi$ [Fig.\,\fref{fig3}{3(b)}]. Such a process is impossible in static systems, as the highest and lowest energy bands cannot touch directly without meeting other existing bands in between. As the eigenspectrum of a Floquet operator is defined on a U(1) loop [see Figs.\,\fref{fig3}{3(h)}--\fref{fig3}{3(k)}], there is no issue for two bands near $E=\pm\pi$ to meet directly and cause a transition. We thus arrive at an anomalous transition between NATPs with a Floquet origin.

\end{multicols}
\vskip 4mm

\fl{3}\centerline{\includegraphics{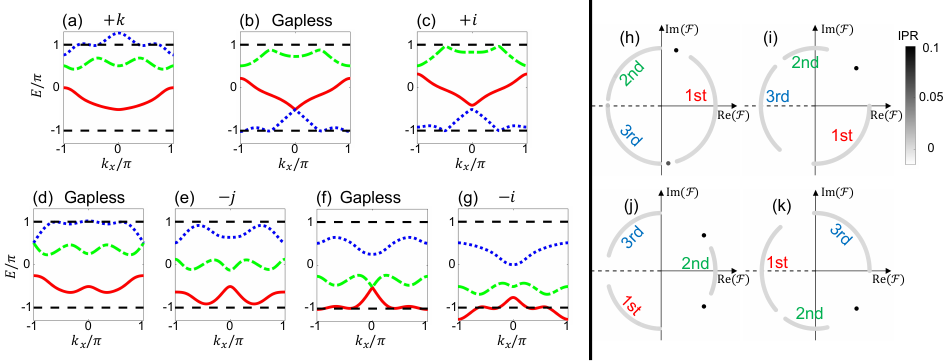}}
\vskip 1mm
\figcaption{15}{3}{(a)--(g) The quasienergy spectra of $U$ in momentum space. The first, second and third bands are marked by the red solid, green dotted and blue dashed lines. (h)--(k) Eigenvalues $\mathcal{F}$ of the Floquet operator on the complex plane. The color bar indicates the inverse participation ratios of different states under OBCs. We fix $v_{CC}=-0.2$ and set $s_{AA}=3$ for (c) and (h); $s_{AA}=2.685$ for (b); $s_{AA}=2$ for (a) and (i); $s_{AA}=1.2$ for (d); $s_{AA}=0$ for (e) and (j); $s_{AA}=-1.2$ for (f); and $s_{AA}=-2$ for (g) and (k). Other parameters are the same as those of Fig.\,\fref{fig2}2. The number of unit cells is $N=500$.}

\begin{multicols}{2}
\baselineskip=12pt plus.2pt minus.2pt

Finally, the usual transition path among NATPs in static systems could also appear in Floquet settings. For example, by decreasing $s_{AA}$ from $s_{AA}=2$ along the red dashed line in Fig.\,\fref{fig2}2, we can encounter two transitions with changes in quaternion charge $k\rightarrow-j\rightarrow-i$. The first (second) transition is accompanied by the vanishing of gap function $\Delta_{23}$ ($\Delta_{12}$) at $k_{x}=\pi$ ($k_{x}=0$), as shown by the spectrum in Fig.\,\fref{fig3}{3(d)} [Fig.\,\fref{fig3}{3(f)}]. A simple mechanism leading to the enrichment of NATPs and phase transitions upon Floquet driving is as follows. In a static setup, the energies have an intrinsic order. An $N$-band static system has at most $(N-1)$ tunable gaps. The lowest band could never meet the highest one without touching the other bands between them. In a Floquet setup, the quasienergies are defined mod $2\pi$ and have no intrinsic order. An $N$-band Floquet system could then have at most $N$ tunable gaps. The two bands beside $E=\pm\pi$ may directly meet or coil around the quasienergy BZ without affecting other bands. Periodic drivings thus create a shortcut to topological transitions, thereby generating NATPs unique to Floquet systems.

{\it 4. Bulk-Edge Correspondence.}
Under open boundary conditions (OBCs), our system has the Floquet operator $U={\rm e}^{-{\rm i}{H}_{2}}{\rm e}^{-{\rm i}{H}_{1}}$, where
\begin{align}
	{H}_{1}=&\sum_{n=1}^{N}\sum_{X={\rm A,B,C}}s_{XX}c_{X,n}^{\dagger}c_{X,n},\\
	{H}_{2}=&\sum_{n=1}^{N-1}\sum_{X,Y={\rm A,B,C}}(v_{XY}c_{X,n}^{\dagger}c_{Y,n+1}+{\rm h.c.}).
\end{align}
The spectrum and eigenstates of $U$ are found by solving the equation $U\ket{\Psi}=\mathcal{F}\ket{\Psi}$, where $\mathcal{F}\equiv\cos E-{\rm i} \sin E$ is the eigenvalue of $U$. We distinguish localized edge states and extended bulk states by their inverse participation ratios (IPRs) (see Sec.\,IV of the SM for details).

Typical spectra of our system under OBCs are shown in Figs.\,\fref{fig3}{3(h)}--\fref{fig3}{3(k)} on the unit circle $|{\cal F}|=1$ of the complex plane. Each black dot contains a pair of degenerate edge modes and each gray bar denotes a bulk Floquet band. System parameters used in Figs.\,\fref{fig3}{3(h)}, \fref{fig3}{3(i)}, \fref{fig3}{3(j)}, and \fref{fig3}{3(k)} are the same as those of Figs.\,\fref{fig3}{3(c)}, \fref{fig3}{3(a)}, \fref{fig3}{3(e)}, and \fref{fig3}{3(g)}, so that they share the same NATCs $i$, $k$, $-j$ and $-i$, respectively. In the case with charge $k$ ($-i$), topological edge modes are found in the gap $\Delta_{12}$ ($\Delta_{23}$) of Fig.\,\fref{fig3}{3(i)} [Fig.\,\fref{fig3}{3(k)}], while the case with charge $-j$ has two pairs of edge modes in both the gaps $\Delta_{12}$ and $\Delta_{23}$ of Fig.\,\fref{fig3}{3(j)} when OBCs are taken. These cases together with their bulk-edge correspondence are the same as NATPs in 1D static systems with ${\cal PT}$-symmetry. The case with charge $i$ in Fig.\,\fref{fig3}{3(h)}, which belongs to the same conjugacy class as charge $-i$, shows an abnormal edge-state configuration in comparison to static situations; i.e., we find two pairs of edge modes separately in the gaps $\Delta_{12}$ and $\Delta_{13}$. As the gap $\Delta_{13}$ is unique to driven systems, the anomalous edge modes observed therein have a Floquet origin. They emerge following a transition path of charge $k\rightarrow i$ via closing and reopening the gap $\Delta_{13}$ without affecting the second band, which is not available in static cases.

Generally, due to the driving-induced quasienergy gap $\Delta_{13}$, the charges $\{\pm i,\pm j,\pm k\}$ of three conjugacy classes could each generate two possible edge-state configurations, with one of them peculiar to Floquet systems. For the $\pm i$ class, edge states can appear in the single gap $\Delta_{23}$ or the two gaps $\Delta_{12}$ and $\Delta_{13}$. For the $\pm j$ class, edge states can appear in the two gaps $\Delta_{12}$ and $\Delta_{23}$ or the single gap $\Delta_{13}$. For the $\pm k$ class, edge states can appear in the single gap $\Delta_{12}$ or the two gaps $\Delta_{13}$ and $\Delta_{23}$. There is thus an intrinsic ``multifold'' relation between bulk NATCs and edge states, which is unique to Floquet systems. The normal and anomalous edge modes for all these classes are sketched in Fig.\,\fref{fig1}{1}. To break the ``multifold'' map from quaternion charges to Floquet edge modes into a one-to-one bulk-edge correspondence, we need to preselect a base point in the parameter space (the green star in Fig.\,\fref{fig2}2 for our case) with a gapped spectrum and fix the order of quasienergy bands there. NATCs throughout the phase diagram can then be uniquely defined. With the known configuration of edge modes at the base point, we can consider any ray from the base point to a different phase, and identify which gap is closed when the critical point is passed along this parameter trajectory. This message, plus the NATCs at two sides of the critical point determine the edge states for the phase entered after the transition uniquely. We elaborate more on this point in Secs.~III and IV of the SM.

{\it 5. Discussion.}
Besides the change from NATCs $k$ to $i$ via a critical point, the transition of charges $j$ (with edge states in the gap $\Delta_{13}$) to $1$ can also be achieved by closing and reopening the quasienergy gap $\Delta_{13}$, making it unique to Floquet systems. Extra subtlety exists in the bulk-edge correspondence of quaternion charge $-1$. In Sec.\,IV of the SM, we clarify the relation between NATCs and edge states via two additional approaches. The first extends $\cos k_{x}$ and $\sin k_{x}$ in the Floquet operator to a parameter space $(k_{1},k_{2})\in\mathbb{R}^{2}$, and relates the edge modes to the locations of bulk Dirac cones of the extended Floquet spectra. The second concerns domain-wall states at the interface of Floquet chains with distinct NATCs. Both approaches yield results that are consistent with our main conclusions. In summary, Floquet driving can not only enrich NATPs and phase transitions, but also create new ones without static analogs. Our theory offers a complete framework to characterize Floquet NATPs in 1D three-band systems. The construction of a dynamical topological invariant\ucite{51} for other non-Abelian Floquet models could be an interesting future direction. As NATPs were already realized in various quantum simulators that are equally suitable for Floquet engineering, our model and its NATPs are within reach of current experiments. For example, the acoustic setup of Ref.\,\cite{63} realized a periodically quenched Floquet lattice. As we only need local quenches, adding an extra leg to the model allows the realization of our three-band lattice  (see Sec.\,V of the SM for more details). Floquet NATPs might then be detected from coherent oscillations of edge modes. Electrical circuits could also be candidates to realize Floquet topological systems, in which continuous time-periodic drives can be implemented.\ucite{64} In Sec.~VI of the SM, we illustrate that our theory of Floquet NATPs is applicable to continuously driven setups. Specially, a harmonic drive is used to open quasienergy gaps and induce Floquet NATPs with quaternion charges $-j$ and $-1$. Their observations could offer further signatures for the presence of Flouqet NATPs.

{\it Acknowledgments.}
This work was supported by the National Natural Science Foundation of China (Grant Nos.~12275260 and  11905211), the Fundamental Research Funds for the Central Universities (Grant No.~202364008), and the Young Talents Project of Ocean University of China.

\end{multicols}

\begin{thebibliography}{99}\footnotesize

\itemsep=-2pt plus.2pt minus.2pt
\bibitem {1} Hasan M Z and Kane C L \href{http://dx.doi.org/10.1103/RevModPhys.82.3045}{2010 {\it Rev. Mod. Phys.} \textbf{82} 3045}
\bibitem {2} Qi X L and Zhang S C \href{http://dx.doi.org/10.1103/RevModPhys.83.1057}{2011 {\it Rev. Mod. Phys.} \textbf{83} 1057}
\bibitem {3} Chiu C K, Teo J C Y, Schnyder A P, and Ryu S \href{http://dx.doi.org/10.1103/RevModPhys.88.035005}{2016 {\it Rev. Mod. Phys.} \textbf{88} 035005}
\bibitem {4} Armitage N P, Mele E J, and Vishwanath A \href{http://dx.doi.org/10.1103/RevModPhys.90.015001}{2018 {\it Rev. Mod. Phys.} \textbf{90} 015001}
\bibitem {5} Zhang X J, Zangeneh-Nejad F, Chen Z G, Lu M H, and Christensen  \href{http://dx.doi.org/10.1038/s41586-023-06163-9}{2023 {\it Nature} \textbf{618} 687}
\bibitem {6} Guo Q H, Jiang T S, Zhang R Y, Zhang L, Zhang Z Q, Yang B, Zhang S, and Chan C T \href{http://dx.doi.org/10.1038/s41586-021-03521-3}{2021 {\it Nature} \textbf{594} 195}
\bibitem {7} Jiang T S, Guo Q H, Zhang R Y, Zhang Z Q, Yang B, and Chan C T \href{http://dx.doi.org/10.1038/s41467-021-26763-1}{2021 {\it Nat. Commun.} \textbf{12} 6471}
\bibitem {8} Jiang T S, Zhang R Y, Guo Q H, Yang B, and Chan C T \href{http://dx.doi.org/10.1103/PhysRevB.106.235428}{2022 {\it Phys. Rev. B} \textbf{106} 235428}
\bibitem {9} Ezawa M \href{http://dx.doi.org/10.1103/PhysRevResearch.3.043006}{2021 {\it Phys. Rev. Res.} \textbf{3} 043006}
\bibitem {10} Liang Y M, Wang R, Yu Z Z, Chen J, Xiao L T, Jia S T, and Zhang L \href{http://dx.doi.org/10.1103/PhysRevB.109.115127}{2024 {\it Phys. Rev. B} \textbf{109} 115127}
\bibitem {11} Wu Q S, Soluyanov A A, and Bzdu\v{s}ek T \href{http://dx.doi.org/10.1126/science.aau8740}{2019 {\it Science} \textbf{365} 1273}
\bibitem {12} Ahn J, Park S, and Yang B J \href{http://dx.doi.org/10.1103/PhysRevX.9.021013}{2019 {\it Phys. Rev. X} \textbf{13} 021013}
\bibitem {13} Bouhon A, Bzdu\v{s}ek T, and Slager R J \href{http://dx.doi.org/10.1103/PhysRevB.102.115135}{2020 {\it Phys. Rev. B} \textbf{102} 115135}
\bibitem {14} Bouhon A, Wu Q S, Slager R J, Weng H M, Yazyev O V, and Bzdu\v{s}ek T \href{http://dx.doi.org/10.1038/s41567-020-0967-9}{2020 {\it Nat. Phys.} \textbf{16} 1137}
\bibitem {15} Park H and Oh S S \href{http://dx.doi.org/10.1088/1367-2630/ac6ca3}{2022 {\it New J. Phys.} \textbf{24} 053042}
\bibitem {16} Jankowski W J, Noormandipour M, Bouhon A, and Slager R J \href{http://dx.doi.org/10.1103/PhysRevB.110.064202}{2024 {\it Phys. Rev. B} \textbf{110} 064202}
\bibitem {17} Breach O, Slager R J, and \"{U}nal F N \href{http://dx.doi.org/10.1103/PhysRevLett.133.093404}{2024 {\it Phys. Rev. Lett.} \textbf{133} 093404}
\bibitem {18} Bouhon A, Zhu Y Q, Slager R J, and Palumbo G \href{http://dx.doi.org/10.1103/PhysRevB.110.195144}{2024 {\it Phys. Rev. B} \textbf{110} 195144}
\bibitem {19} Park S, Hwang Y, Choi H C, and Yang B J \href{http://dx.doi.org/10.1038/s41467-021-27158-y}{2021 {\it Nat. Commun.} \textbf{12} 6781}
\bibitem {20} Peng B, Bouhon A, Slager R J, and Monserrat B \href{http://dx.doi.org/10.1103/PhysRevB.105.085115}{2022 {\it Phys. Rev. B} \textbf{105} 085115}
\bibitem {21} Park H, Wong S, Bouhon A, Slager R J, and Oh S S \href{http://dx.doi.org/10.1103/PhysRevB.105.214108}{2022 {\it Phys. Rev. B} \textbf{105} 214108}
\bibitem {22} Peng B, Bouhon A, Monserrat B, and Slager R J \href{http://dx.doi.org/10.1038/s41467-022-28046-9}{2022 {\it Nat. Commun.} \textbf{13} 423}
\bibitem {23} Lange G F, Bouhon A, Monserrat B, and Slager R J \href{http://dx.doi.org/10.1103/PhysRevB.105.064301}{2022 {\it Phys. Rev. B} \textbf{105} 064301}
\bibitem {24} Sun X C, Wang J B, He C, and Chen Y F \href{http://dx.doi.org/10.1103/PhysRevLett.132.216602}{2024 {\it Phys. Rev. Lett.} \textbf{132} 216602}
\bibitem {25} Bouhon A, Lange G F, and Slager R J \href{http://dx.doi.org/10.1103/PhysRevB.103.245127}{2021 {\it Phys. Rev. B} \textbf{103} 245127}
\bibitem {26} Chen S Y, Bouhon A, Slager R J, and Monserrat B \href{http://dx.doi.org/10.1103/PhysRevB.105.L081117}{2022 {\it Phys. Rev. B} \textbf{105} L081117}
\bibitem {27} \"{U}nal F N, Bouhon A, and Slager R J \href{http://dx.doi.org/10.1103/PhysRevLett.125.053601}{2020 {\it Phys. Rev. Lett.} \textbf{125} 053601}
\bibitem {28} Zhao W D, Yang Y B, Jiang Y, Mao Z C, Guo W X, Qiu L Y, Wang G X, Yao L, He L, Zhou Z C {\it et al.} \href{http://dx.doi.org/10.1038/s42005-022-01001-2}{2022 {\it Commun. Phys.} \textbf{5} 223}
\bibitem {29} Wang Q D, Zhu Y Q, Zhu S L, and Zheng Z \href{http://dx.doi.org/10.1103/PhysRevA.110.023321}{2024 {\it Phys. Rev. A} \textbf{110} 023321}
\bibitem {30} Yan Q C, Wang Z H, Wang D Y, Ma R, Lu C C, Ma G C, Hu X Y, and Gong Q H \href{http://dx.doi.org/10.1364/AOP.494544}{2023 {\it Adv. Opt. Photon.} \textbf{15} 907}
\bibitem {31} Yang Y, Yang B, Ma G C, Li J, Zhang S, and Chan C T \href{http://dx.doi.org/10.1126/science.adf9621}{2024 {\it Science} \textbf{383} eadf9621}
\bibitem {32} Cayssol J, D\'{o}ra B, Simon F, and Moessner R \href{http://dx.doi.org/10.1002/pssr.v7.1/2}{2013 {\it Phys. Status Solidi RRL} \textbf{7} 101}
\bibitem {33} Bukov M, D'Alessio L, and Polkovnikov A \href{http://dx.doi.org/10.1080/00018732.2015.1055918}{2015 {\it Adv. Phys.} \textbf{64} 139}
\bibitem {34} Eckardt A \href{http://dx.doi.org/10.1103/RevModPhys.89.011004}{2017 {\it Rev. Mod. Phys.} \textbf{89} 011004}
\bibitem {35} Oka T and Kitamura S \href{http://dx.doi.org/10.1146/annurev-conmatphys-031218-013423}{2019 {\it Annu. Rev. Condens. Matter Phys.} \textbf{10} 387}
\bibitem {36} Rudner M S and Lindner N H \href{http://dx.doi.org/10.1038/s42254-020-0170-z}{2020 {\it Nat. Rev. Phys.} \textbf{2} 229}
\bibitem {37} Harper F, Roy R, Rudner M S, and Sondhi S L \href{http://dx.doi.org/10.1146/annurev-conmatphys-031218-013721}{2020 {\it Annu. Rev. Condens. Matter Phys.} \textbf{11} 345}
\bibitem {38} Bandyopadhyay S, Bhattacharjee S, and Sen D \href{http://dx.doi.org/10.1088/1361-648X/ac1151}{2021 {\it J. Phys. Condens. Matter} \textbf{33} 393001}
\bibitem {39} de la Torre A, Kennes D M, Claassen M, Gerber S, McIver J W, and Sentef M A \href{http://dx.doi.org/10.1103/RevModPhys.93.041002}{2021 {\it Rev. Mod. Phys.} \textbf{93} 041002}
\bibitem {40} Zhou L W and Zhang D J \href{http://dx.doi.org/10.3390/e25101401}{2023 {\it Entropy} \textbf{25} 1401}
\bibitem {41} Zhan F Y, Chen R, Ning Z, Ma D S, Wang Z M, Xu D H, and Wang R \href{http://dx.doi.org/10.1007/s44214-024-00067-z}{2024 {\it Quantum Front.} \textbf{3} 21}
\bibitem {42} Oka T and Aoki H \href{http://dx.doi.org/10.1103/PhysRevB.79.081406}{2009 {\it Phys. Rev. B} \textbf{79} 081406}
\bibitem {43} Lindner N H, Refael G, and Galitski V \href{http://dx.doi.org/10.1038/nphys1926}{2011 {\it Nat. Phys.} \textbf{7} 490}
\bibitem {44} Rechtsman M C, Zeuner J M, Plotnik Y, Lumer Y, Podolsky D, Dreisow F, Nolte S, Segev M, and Szameit A \href{http://dx.doi.org/10.1038/nature12066}{2013 {\it Nature} \textbf{496} 196}
\bibitem {45} Jotzu G, Messer M, Desbuquois R, Lebrat M, Uehlinger T, Greif D, and Esslinger T \href{http://dx.doi.org/10.1038/nature13915}{2014 {\it Nature} \textbf{515} 237}
\bibitem {46} Ho D Y H and Gong J B \href{http://dx.doi.org/10.1103/PhysRevLett.109.010601}{2012 {\it Phys. Rev. Lett.} \textbf{109} 010601}
\bibitem {47} Tong Q J, An J H, Gong J B, Luo H G, and Oh C H \href{http://dx.doi.org/10.1103/PhysRevB.87.201109}{2013 {\it Phys. Rev. B} \textbf{87} 201109}
\bibitem {48} Xiong T S, Gong J B, and An J H \href{http://dx.doi.org/10.1103/PhysRevB.93.184306}{2016 {\it Phys. Rev. B} \textbf{93} 184306}
\bibitem {49} Yang K, Xu S Y, Zhou L W, Zhao Z Y, Xie T Y, Ding Z, Ma W C, Gong J B, Shi F Z, and Du J F \href{http://dx.doi.org/10.1103/PhysRevB.106.184106}{2022 {\it Phys. Rev. B} \textbf{106} 184106}
\bibitem {50} Jiang L, Kitagawa T, Alicea J, Akhmerov A R, Pekker D, Refael G, Cirac J I, Demler E, Lukin M D, and Zoller P \href{http://dx.doi.org/10.1103/PhysRevLett.106.220402}{2011 {\it Phys. Rev. Lett.} \textbf{106} 220402}
\bibitem {51} Rudner M S, Lindner N H, Berg E, and Levin M \href{http://dx.doi.org/10.1103/PhysRevX.3.031005}{2013 {\it Phys. Rev. X} \textbf{3} 031005}
\bibitem {52} Zhou L W, Chen C, and Gong J B \href{http://dx.doi.org/10.1103/PhysRevB.94.075443}{2016 {\it Phys. Rev. B} \textbf{94} 075443}
\bibitem {53} Rodriguez-Vega M and Seradjeh B \href{http://dx.doi.org/10.1103/PhysRevLett.121.036402}{2018 {\it Phys. Rev. Lett.} \textbf{121} 036402}
\bibitem {54} Kitagawa T, Berg E, Rudner M, and Demler E \href{http://dx.doi.org/10.1103/PhysRevB.82.235114}{2010 {\it Phys. Rev. B} \textbf{82} 235114}
\bibitem {55} Nathan F and Rudner M S \href{http://dx.doi.org/10.1088/1367-2630/17/12/125014}{2015 {\it New J. Phys.} \textbf{17} 125014}
\bibitem {56} Yao S Y, Yan Z B, and Wang Z \href{http://dx.doi.org/10.1103/PhysRevB.96.195303}{2017 {\it Phys. Rev. B} \textbf{96} 195303}
\bibitem {57} Roy R and Harper F \href{http://dx.doi.org/10.1103/PhysRevB.96.155118}{2017 {\it Phys. Rev. B} \textbf{96} 155118}
\bibitem {58} Slager R J, Bouhon A, and \"{U}nal F N \href{http://dx.doi.org/10.1038/s41467-024-45302-2}{2024 {\it Nat. Commun.} \textbf{15} 1144}
\bibitem {59} Li T Y and Hu H P \href{http://dx.doi.org/10.1038/s41467-023-42139-z}{2023 {\it Nat. Commun.} \textbf{14} 6418}
\bibitem {60} Karle V, Lemeshko M, Bouhon A, Slager R J, and \"{U}nal F N  \href{https://arxiv.org/abs/2408.16848}{2024 {arXiv:2408.16848} [quant-ph]}
\bibitem {61} Slager R J, Bouhon A, and \"{U}nal F N  \href{https://arxiv.org/abs/2310.12782}{2023 {arXiv:2310.12782} [cond-mat.mes-hall]}
\bibitem {62} Asb\'{o}th J K and Obuse H \href{http://dx.doi.org/10.1103/PhysRevB.88.121406}{2013 {\it Phys. Rev. B} \textbf{88} 121406}
\bibitem {63} Cheng Z Y, Bomantara R W, Xue H R, Zhu W W, Gong J B, and Zhang B L \href{http://dx.doi.org/10.1103/PhysRevLett.129.254301}{2022 {\it Phys. Rev. Lett.} \textbf{129} 254301}
\bibitem {64} Zhang W X, Cao W H, Qian L, Yuan H, and Zhang X D \href{http://dx.doi.org/10.1038/s41467-024-55425-1}{2025 {\it Nat. Commun.} \textbf{16} 198}

\end{thebibliography}
\end{document}